# Transverse angular momentum in topological photonic crystals

Wei-Min Deng, Xiao-Dong Chen, Fu-Li Zhao and Jian-Wen Dong

School of Physics & State Key Laboratory of Optoelectronic Materials and Technologies, Sun Yat-sen University, Guangzhou 510275, China

E-mail: dongjwen@mail.sysu.edu.cn

**Abstract**

Engineering local angular momentum of structured light fields in real space enables unprecedented applications in many fields, in particular for the realization of unidirectional robust transport in topological photonic crystals with non-trivial Berry vortex in momentum space. Here, we show transverse angular momentum modes in silicon topological photonic crystals when considering transverse electric polarization. Excited by a chiral external source with either transverse spin or orbital angular momentum, robust light flow propagating along opposite directions was observed in several kinds of sharp-turn interfaces between two topologically-distinct silicon photonic crystals. A transverse orbital angular momentum mode with alternating-sign topological charge was found at the boundary of such two photonic crystals. In addition, we also found that unidirectional transport is robust to the working frequency even when the ring-size or location of pseudo-spin source varies in a certain range, leading to the superiority of broadband photonic device. These findings enable for making use of transverse angular momentum, a kind of degree of freedom, to achieve unidirectional robust transport in telecom region and other potential applications in integrated photonic circuits such as on-chip robust delay line.

# 1. Introduction

Structured light fields in nanoscale have attracted unprecedented attention and have brought many applicaitons, such as nanometric optical tweezer [1], atomic manipulation [2], quantum communication and information [3]. As structured light fields can carry angular momentum, e.g., spin angular momentum (SAM) associated with circular polarization and orbital angular momentum (OAM) derived from phase structure of light, one may explore such two important degrees of freedom to achieve unique electromagnetic behaviors and light-matter interaction between structured light fields and nanostructures. One particular example is to employ angular momentum of structured light fields to excite the valley chiral bulk state in valley photonic crystals [4, 5]. Another issue is to utilize local chirality in near fields of nanophotonic structure to realize directional photon emission [6, 7]. The discovery of such chiral light-matter interaciton brings about the research field of chiral quantum optics [8].

Topological photonics is one of fantastic topics in recent years, with certain superiority to its predecessor in condensed matter physics, e.g., easy sample prepration and room temperature characterization. Topology degree of freedom in photonics provides a new point of view to mold robust unidirectional flow of light in various systems [9-29]. However, it brings about a tricky problem in experiment such that how to construct proper pseudo-spin source in different types of photonic systems with time-reversal symmetry protection. In previous literatures, several schemes have been proposed to achieve unidirectional transport without backscattering, but they are tough and impractical in telecom region [13-17]. Compared with pseudo-spin, a true spin, i.e. electric circular polarization, is easier to construct in telecom region. Due to the fact that optical circularly-polarized source is naturally in terms of electric fields, the key point for the excitation of the unidirectional edge state by a true spin is to have a topological photonic crystal of transverse-electric (TE) polarization and to investigate the local chirality of structured light fields inside. Considering that electric circularly-polarized light may carry orbital angular momentum, a true spin carrying spin and orbital angular momenta will be native for achieving topological functionality on chip-size systems.

In this work, we studied transverse spin and orbital angular momenta in silicon topological photonic crystals when considering TE polarization. By plotting polarization ellipse of electric field near the boundary of topological photonic crystals, we found several circular polarization points, of which the handedness is locked to the light propagating direction. Meanwhile, we observed phase vortexes with the handedness locking to the light direction. By exciting transverse spin or orbital angular momentum mode with different chirality, robust light flow propagating along towards opposite direction was achieved in several kinds of sharp-turn interfaces between two topologically-distinct silicon photonic crystals. Specially, there was a transverse orbital angular momentum (TOAM) mode with alternating-sign topological charge in the phase distribution of $E_y$ component. Besides, we also found the unidirectional light flow is robust to the working frequency even when the ring-size or the position of pseudo-spin source are changed. By taking advantage of transverse angular momentum, the above findings enable to realize robust and unidirectional light flow in telecom region and bring potential applications in integrated photonic circuits such as on-chip robust delay line.

## 2. Topological photonic crystals

Here, we adopted the $\vec{k} \cdot \vec{p}$ method in [30] to derive the Hamiltonian of photonic crystals. Similar Hamiltonian has been studied in TM polarization [15], while we consider TE polarization in this paper. So we have non-zero $E_x$, $E_y$ and $H_z$ components. For TE polarized states in two-dimensional photonic crystals, the eigenvalue equation is expressed as

$$-\nabla \cdot [\frac{1}{\varepsilon_r(\vec{r})}\nabla H_z] = \frac{\omega^2}{c^2}H_z. \tag{1}$$

After applying Bloch theory, we can express $H_z$ as $H_z(\vec{r}) = e^{i\vec{k}\cdot\vec{r}}u(\vec{r})$, where $u(\vec{r})$ is the periodic function. Then Eq. (1) turns to

$$(\hat{H}_0 + \hat{H}_{pert})u(\vec{r}) = Eu(\vec{r}), \tag{2}$$

where $\hat{H}_0 = -\frac{1}{\varepsilon_r(\vec{r})}\nabla^2 - [(\frac{\partial}{\partial x}\frac{1}{\varepsilon_r(\vec{r})})\cdot\frac{\partial}{\partial x} + (\frac{\partial}{\partial y}\frac{1}{\varepsilon_r(\vec{r})})\cdot\frac{\partial}{\partial y}]$, $\hat{H}_{pert} = -\frac{2i}{\varepsilon_r(\vec{r})}\vec{k}\cdot\nabla - i\vec{k}\cdot\nabla\frac{1}{\varepsilon_r(\vec{r})} + \frac{k^2}{\varepsilon_r(\vec{r})}$ and $E = \omega^2/c^2$. $\hat{H}_0$ is the unperturbed part and $\hat{H}_{pert}$ is the perturbation part away from $\Gamma$ point. Assume there are two pairs of degenerate eigenfunctions at $\Gamma$ point, with the eigenvalues to be $E_1 = E_2$ and $E_3 = E_4$. The parity of each eigenfunction is denoted as $f_1 = p_x$, $f_2 = p_y$, $f_3 = d_{x^2-y^2}$, $f_4 = d_{2xy}$. Then, on the basis of $[p_x, p_y, d_{x^2-y^2}, d_{2xy}]$, if we consider the perturbation near $\Gamma$ point, the Hamiltonian has a $4\times4$ matrix form

$$\mathbf{H} = \begin{pmatrix} \mathbf{H_{pp}} & \mathbf{H_{pd}} \\ \mathbf{H_{pd}^\dagger} & \mathbf{H_{dd}} \end{pmatrix}, \tag{3}$$

where each component is a $2\times2$ matrix, and $\dagger$ is the conjugate transpose operator. We derived the matrix elements of $\mathbf{H_{pp}}$ by applying second-order degenerate $\vec{k}\cdot\vec{p}$ perturbation theory to the degenerate states $f_1$ and $f_2$ [30]. Each component in $\mathbf{H_{pp}}$ is expressed as

$$\mathrm{H}_{pp}^{mn} = E_m\delta_{mn} + H'_{mn} + \sum_{\alpha=3,4}\frac{H'_{m\alpha}H'_{\alpha n}}{(E_m - E_\alpha)}, \tag{4}$$

where $m = 1, 2$, $n = 1, 2$ and $H'_{mn} = \langle f_m|\hat{H}_{pert}|f_n\rangle$. For example, $\mathrm{H}_{pp}^{11}$ can be written as

$$\mathrm{H}_{pp}^{11} = E_1 + \langle f_1|\hat{H}_{pert}|f_1\rangle + \sum_{\alpha=3,4}\frac{\langle f_1|\hat{H}_{pert}|f_\alpha\rangle\langle f_\alpha|\hat{H}_{pert}|f_1\rangle}{E_1 - E_\alpha} \approx E_1 + q_1k^2 + Gk_x^2 + Fk_y^2, \tag{5}$$

where $q_1 = \langle f_1 | 1/\varepsilon_r(\vec{r}) | f_1 \rangle$, $G = (|\langle p_x | -\frac{2i}{\varepsilon_r(\vec{r})} \cdot \frac{\partial}{\partial x} | d_{x^2-y^2} \rangle|^2 - |\langle p_x | -i(\frac{\partial}{\partial x} \frac{1}{\varepsilon_r(\vec{r})}) | d_{x^2-y^2} \rangle|^2)/(E_1 - E_3)$ and

$F = (|\langle p_x | -\frac{2i}{\varepsilon_r(\vec{r})} \cdot \frac{\partial}{\partial y} | d_{2xy} \rangle|^2 - |\langle p_x | -i(\frac{\partial}{\partial y} \frac{1}{\varepsilon_r(\vec{r})}) | d_{2xy} \rangle|^2)/(E_1 - E_3)$. According to Eq. (4), we can obtain other components in $\mathbf{H_{pp}}$. Finally, we can express $\mathbf{H_{pp}}$ as

$$\mathbf{H_{pp}} = \begin{pmatrix} E_1 + q_1 k^2 + G k_x^2 + F k_y^2 & N k_x k_y \\ N k_x k_y & E_1 + q_2 k^2 + F k_x^2 + G k_y^2 \end{pmatrix}, \quad (6)$$

where $N \approx F + G$ and $q_2 = \langle f_2 | 1/\varepsilon_r(\vec{r}) | f_2 \rangle$. In the same way, by applying second-order degenerate $\vec{k} \cdot \vec{p}$ perturbation theory to the degenerate states $f_3$ and $f_4$, $\mathbf{H_{dd}}$ can also be obtained and is expressed as

$$\mathbf{H_{dd}} = \begin{pmatrix} E_3 + q_3 k^2 - G k_x^2 - F k_y^2 & -N k_x k_y \\ -N k_x k_y & E_3 + q_4 k^2 - F k_x^2 - G k_y^2 \end{pmatrix}. \quad (7)$$

For non-diagonal terms $\mathbf{H_{pd}}$ and $\mathbf{H_{pd}^\dagger}$ in Eq. (3), these components can be expressed as

$\mathrm{H}_{pd}^{11} = \langle f_1 | \hat{H}_{pert} | f_3 \rangle = A k_x$, $\mathrm{H}_{pd}^{12} = \langle f_1 | \hat{H}_{pert} | f_4 \rangle \approx A k_y$, $\mathrm{H}_{pd}^{21} = \langle f_2 | \hat{H}_{pert} | f_3 \rangle \approx -A k_y$,

$\mathrm{H}_{pd}^{22} = \langle f_2 | \hat{H}_{pert} | f_4 \rangle \approx A k_x$. The Hamiltonian $\mathbf{H}$ is thus given by

$$\mathbf{H} = \begin{pmatrix} E_1 + q_1 k^2 + G k_x^2 + F k_y^2 & N k_x k_y & A k_x & A k_y \\ N k_x k_y & E_1 + q_2 k^2 + F k_x^2 + G k_y^2 & -A k_y & A k_x \\ A^* k_x & -A^* k_y & E_3 + q_3 k^2 - G k_x^2 - F k_y^2 & -N k_x k_y \\ A^* k_y & A^* k_x & -N k_x k_y & E_3 + q_4 k^2 - F k_x^2 - G k_y^2 \end{pmatrix}, \quad (8)$$

where $q_j = \langle f_j | 1/\varepsilon_r(\vec{r}) | f_j \rangle, (j = 1, 2, 3, 4)$. Considering that $q_j (j = 1, 2, 3, 4)$ is much smaller than $F$ and $G$ when the degeneration between $[p_x, p_y]$ and $[d_{x^2-y^2}, d_{2xy}]$ is slightly broken, we can neglect the $q_j (j = 1, 2, 3, 4)$ in Eq. (8). Moreover, if we neglect the second order off-diagonal terms and shift the zero-energy point to $(E_1 + E_3)/2$, the Hamiltonian will be block diagonalized under a new basis of $[p_+, d_+, p_-, d_-]$, where $p_\pm = (p_x \pm i p_y)/\sqrt{2}$ and $d_\pm = (d_{x^2-y^2} \pm i d_{2xy})/\sqrt{2}$, yielding

$$\mathbf{H_1} = \begin{pmatrix} M + Bk^2 & Ak_- & 0 & 0 \\ A^*k_+ & -M - Bk^2 & 0 & 0 \\ 0 & 0 & M + Bk^2 & Ak_+ \\ 0 & 0 & A^*k_- & -M - Bk^2 \end{pmatrix}, \quad (9)$$

where $B = (F + G)/2$, $M = (E_1 - E_3)/2$ and $k_\pm = k_x \pm ik_y$. It is interesting to see that the form of Eq. (9) is similar to the Hamiltonian of Bernevig-Hughes-Zhang model, implying a topological band gap if the band inversion occurs [31]. Similar to the Bernevig-Hughes-Zhang model, we can evaluate the spin Chern number of the topological photonic crystals as [32]

$$C_\pm = \pm \frac{1}{2}[\text{sgn}(M) + \text{sgn}(-B)]. \quad (10)$$

In Eq. (10), parameter $B$ is typically negative [33], thus the spin Chern number will be non-zero if the sign of parameter $M$ is positive, namely the frequency of $p$ states is higher than that of $d$ states.

To achieve the non-zero spin Chern number, we have to firstly obtain the four specific eigenstates $[p_x, p_y, d_{x^2-y^2}, d_{2xy}]$. Because the on-demand degenerate states are the basis functions of two two-dimensional irreducible representations in $C_{6v}$ point group, it is straightforward to choose the hexagonal lattice. Note that the hexagonal lattice with a simple unit cell possesses a Dirac cone at K point, while the selection of compound unit cell can result in band folding, ensuring the emergence of two sets of degenerate points at $\Gamma$ point. In this way, a silicon photonic crystal is designed in a compound unit cell [figure 1(a)] consisting of two types of dielectric rods (blue) embedded in air background. The two rods have the same permittivity of $\varepsilon_r = 12.08$, while the radii of the center and corner rods are $r_1$ and $r_2$, respectively. The distance between the nearest-neighboring rods is $b = a/\sqrt{3}$, where $a$ is the lattice constant. When $r_1 = r_2 = 0.3b$, such silicon photonic crystal has double Dirac cones at $\Gamma$ point at the degenerate frequency of 0.799 $c/a$ [figure 1(c)]. To break the cone and open a topological gap, we should let the frequency of the $p$ states higher than that of the $d$ states, and it can be achieved when $(r_1, r_2) = (0, 0.33b)$ [figure 1(d)]. The highlighted *p* and *d* states in figure 1(d) are identified by analyzing the parity of $H_z$ field patterns. On the contrary, a trivial gap opens when $(r_1, r_2) = (0.35b, 0.26b)$, as the *p* states locate below *d* states in figure 1(b). Note that these two topologically-distinct gaps share a common frequency range, resulting in a fantastic characteristic of robust edge state at the boundary of such two silicon photonic crystals [figure 2(a)].

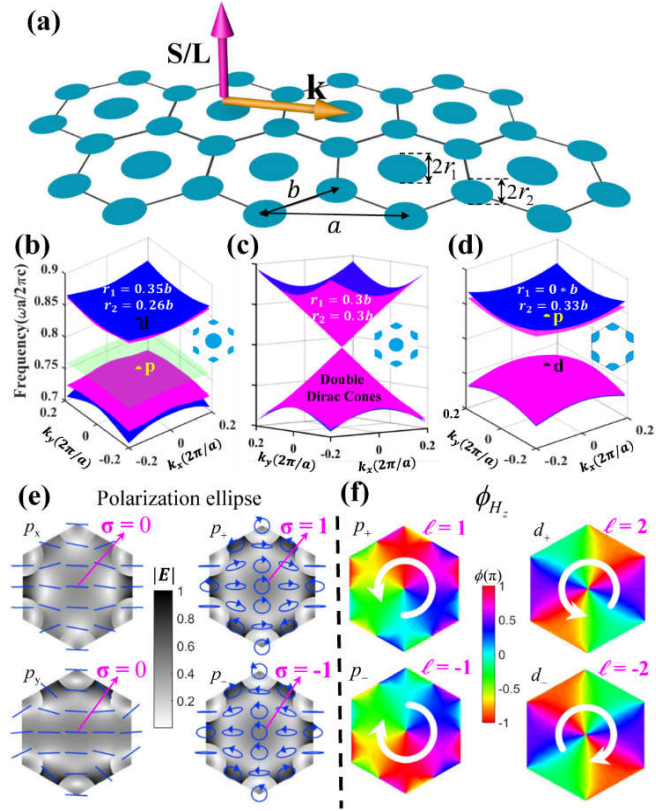

**Figure 1.** Transverse spin and orbital angular momenta in silicon photonic crystals when considering transverse-electric polarization. (a) Hexagonal photonic crystal with a compound unit cell consisting of two dielectric rods (blue) in the air background. The permittivity of the two rods are $\varepsilon_r = 12.08$, while the radii of the center and corner rods are denoted as $r_1$ and $r_2$, respectively. The distance between the two rods is $b = a/\sqrt{3}$, where *a* is the lattice constant. The yellow and pink arrows in (a) indicate the direction of wave vector and transversely local spin/orbital angular momenta, respectively. (b)-(d) Photonic band structures with various $(r_1, r_2)$ configurations, (b) trivial gap and $(0.35b, 0.26b)$, (c) double Dirac cones and $(0.3b, 0.3b)$, (d) nontrivial gap and $(0, 0.33b)$. Note that the double Dirac cones in (c) results from band folding from zone boundary as the unit cell is chosen to the supercell. (e) Transverse spin and (f) transverse orbital angular momentum at $\Gamma$ point in topological photonic crystal with the configuration in (d). Note that such transverse angular momentum feature also exists in trivial crystal in (b). Here, $\sigma = \pm 1$ represents RCP/LCP electric field, and $l = \pm 1$ labels the spiral phase distributions of magnetic field.

## 2. Transverse angular momentum

In this section, we will demonstrate the transverse spin and orbital angular momenta of topological photonic crystals. Here, the term 'transverse' indicates the direction of angular momentum (pink arrow) is orthogonal to the propagating direction (brown arrow) [figure 1(a)]. In general, one can recombine the in-plane electric fields of the $p_x$ and $p_y$ states when considering the $\pi/2$ rotation in between. For example, we can have an expression with the form of $\frac{i}{\omega\varepsilon_r(\vec{r})}\nabla\times((p_x\pm ip_y)\hat{z})=(E_{1x}\mp iE_{1y})(\hat{x}\pm i\hat{y})$ at the unit cell center, where $E_{1x}$ and $E_{1y}$ are the in-plane electric fields of $p_x$ state, showing that the pseudo-spin states $p_\pm(=(p_x\pm ip_y)/\sqrt{2})$ are of an intrinsic right/left-handed circular polarization. This is verified in local polarization ellipse map by retrieving the in-plane electric fields of the eigenfunction at $\Gamma$ point. It is clear to see in figure 1(e), that the local polarization of the $p_x$ and $p_y$ states is always linear both at and beyond the center of unit cell (labelled by blue segments in left column, denoted as $\sigma=0$), while the local polarization ellipse of the $p_\pm$ states is indeed electrically circular polarized at the unit cell center (labeled by blue circle ring in right column, denoted as $\sigma=\pm 1$). Here, $\sigma=1$ is right-handed circular polarization (RCP) and $\sigma=-1$ is left-handed circular polarization (LCP). Note that the local polarization in most of the unit cell is circular or ellipse in the $p_\pm$ states. In other words, the transverse spin angular momentum (TSAM) is intrinsic and general in topological photonic crystals.

Note that the $p_x$ and $p_y$ states share the same parities as the two Hermite-Gaussian modes ($HG_{10}$ and $HG_{01}$), and the latters directly connect to Laguerre-Gaussian modes carrying orbital angular momentum [34], with the form of $LG_{01}=(HG_{10}+iHG_{01})/\sqrt{2}$ and $LG_{10}=(HG_{10}-iHG_{01})/\sqrt{2}$. So it is straightforward to expect TOAM in the $p_\pm$ states. This is demonstrated by the emergence of phase vortex with the topological charge of $\pm 1$, as shown in the left column of figure 1(f). The spiral phase distribution is obvious near the center of the unit cell, but with opposite chirality in the $p_+$ and $p_-$ states. In addition, it is provable that the $d_\pm$ states also carry intrinsic TOAM and the topological charge of phase vortexes is $\pm 2$. This is also consistent with the calculated phase pattern in right column of figure 1(f).

## 3. Robust unidirectional propagation by excitation of transverse angular momentum modes

Transverse spin/orbital angular momentum edge mode is crucial for the excitation of robust unidirectional propagation in topological photonic crystals. Figure 2(a) shows the dispersion relation of the TOAM edge mode, spanning the whole photonic band gap from the frequency of 0.769 *c/a* to 0.837 *c/a*, except a minigap at $\Gamma$ point due to the broken $C_{6v}$ symmetry at the interface. The mode profile (with the frequency of 0.823 *c/a*) near the boundary between two topologically-distinct photonic crystals is also illustrated in figures 2(b)-2(d). For the pseudo-spin up edge state [blue curve in figure 2(a)], it is clear to observe a $H_z$ phase vortex with a clockwise gradient change ($l=-1$) [white arrow in the second panel of figure 2(c)] and zero $H_z$ intensity at the center [the leftmost panel in figure 2(c)] due to the undefined phase at this singularity point. The pseudo-spin down edge state [red cruve in figure 2(a)] which is the time-reversal counterpart of pseudo-spin up state has a phase vortex with $l=1$, as marked by the counterclockwise white arrow in figure 2(d). In order to quantify such phase vortex in the whole forbidden gap, a vortex factor $Q = \int_{\text{square}} (\nabla \times <\vec{S}>) \cdot d\vec{A}$ is used followed by Ref. [35]. Here, $<\vec{S}>$ is the time-averaged Poynting vector, and the integration area is an inscribed square of the dielectric rod possessing phase vortex, i.e., local TOAM. The pseudo-spin edge dispersion in figure 2(a) is then colored by the value of vortex factor. The vortex factor magnitude, proportional to the total storing energy of TOAM in each pseudo-spin edge state, reaches a maximum value inside the bandgap while goes to minmum near either Brillouin zone center or upper bandgap boundary due to spin (vortex) mixture. The sign of vortex factor, locking to the direction of local TOAM and the corresponding topological charge in real space, is negative for most of the pseudo-spin up edge states while positive for the time-reversal partner. Note that the group velocity of the pseudo-spin up/down edge state is always positive/negative. One can infer that the sign of vortex factor (also topological charge of TOAM) is locked to the propagation direction of the light flow. In other words, a TOAM external source with topological charge of 1 or -1 can be utilized to excite robust unidirectional edge state at the boundary of topological photonic crystals (will be discussed later).

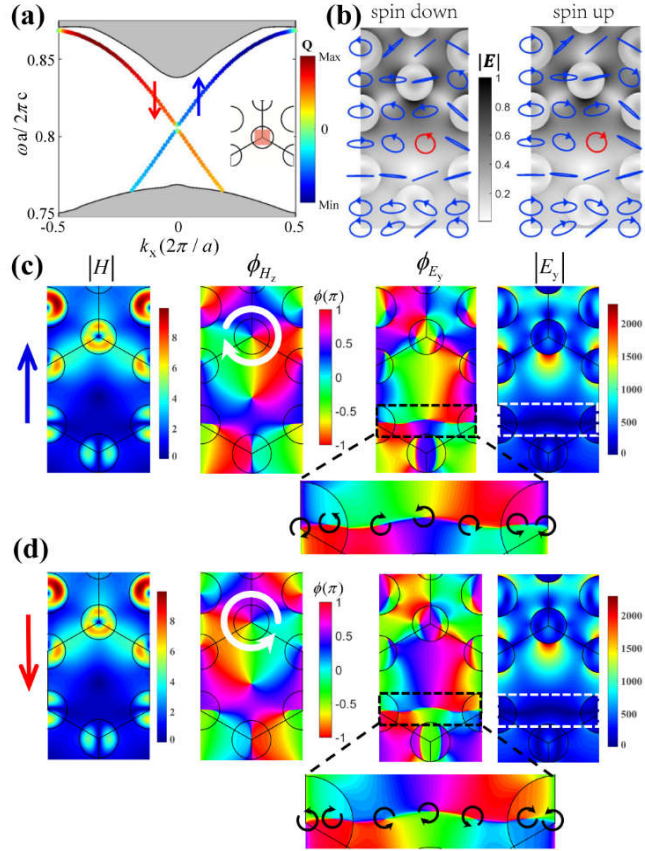

**Figure 2.** Transverse spin and orbital angular momenta at the boundary of topological photonic crystals. (a) Dispersion relation of robust edge state spanning within the whole forbidden gap except near zone center. A vortex factor Q is employed to evaluate the feature of transverse orbital angular momentum in the pseudo-spin up/down (blue/red) edge states. Positive/negative Q indicates anti-clockwise/clockwise phase vortex of $H_z$ field. The magnitude of Q is proportional to the total storing energy. Inset, the red square is the integration area in the Q calculation. (b) Polarization ellipse distribution when $f = 0.823\ c/a$ and $k_x = \mp 0.08 \times 2\pi/a$, verifying the existence of transverse spin angular momentum (TSAM) near the crystal interface. The blue line and circle segments represent various local polarization states at different position. The background is the magnitude of normalized electric field. (c)-(d) Magnitude and phase distribution on $H_z$ and $E_y$ components around the crystal interface for the pseudo-spin edge states. The white and black arrows indicate the direction of phase gradient change. The zoom-in sections show the vortex chain with alternative sign of topological charge.

Alternatively, one can also employ a transverse spin angular momentum (TSAM) source to control the robust unidirectional light flow. Figure 2(b) shows the TSAM mode profile near the edge at the frequency of 0.823 $c/a$. The polarization ellipse of the electric field is obviously different point-by-point, and some of positions have chiral electric fields. For example, at the point highlighted in red circle, it is RCP for the pseudo-spin up edge state while LCP for the pseudo-spin down one. Therefore, a chiral quantum dot placing at the red point may serve as a TSAM source to excite unidirectional edge state, which is similar to Ref. [6] and will be discussed later.

In addtion, the two rightmost panels of figures 2(c)-2(d) show the $E_y$ fields of the pair of the pseudo-spin edge states. A one-dimensional vortex chain can be observed in a near-zero intensity distribution of $E_y$ field (dashed frame). After labeling the direction of phase gradient (black arrows in zoom-in figures), it is further found the alternative sign arragement of the vortex chain, with similar feature to the phase structure of propagating beam whose initial phase is of half-integer topological charge [36]. Note that the rotation directions in each vortex are inverse between the two pseudo-spin edge states due to time-reversal-symmetry protection.

Exciting robust unidirectional light flow at the boudnary of topological photonic crsytals needs a source carrying transverse angular momentum. To mimic a TOAM source, twenty-four equal-amplitude $H_z$ point sources are set isometrically on a circle, namely ring source hereafter. The phase of such ring source increases clockwise ($l=-1$) or anti-clockwise ($l=1$) with linear interval of $\pi/12$. Figure 3 plots the topologically-protected results when the radius of ring source is $R_s = 0.6b$, with the circle center at the white-arrow-surrounding rod in figure 2(c). One can see that the electromagnetic wave propagates in one-way direction with little reflection, e.g., rightward flow by $l=-1$ source excitation in figure 3(a). The excited waves smoothly pass through a series of $60°/120°$ sharp corners at the boundary of topological photonic crystal. Moreover, the unidirectional flow has a broadband feature within the whole topological photonic bandgap, see the green curve of figure 4(a). Such unique behavior also occurs when placing a chiral dipole source (i.e. RCP or LCP) at the position with chiral polarization ellipse [red in figure 2(b)]. Note that the chiral dipole source is used to mimic the TSAM mode, with the phase difference between $E_x$ and $E_y$ of $\pi/2$ ($\sigma=1$) and $-\pi/2$ ($\sigma=-1$), respectively. Note also that we demonstrate a different sharp corner with a $90°$ bend in figure 3(b), in order to illustrate the diverse design proposals of robust unidirectional flow in the future chip-size device.

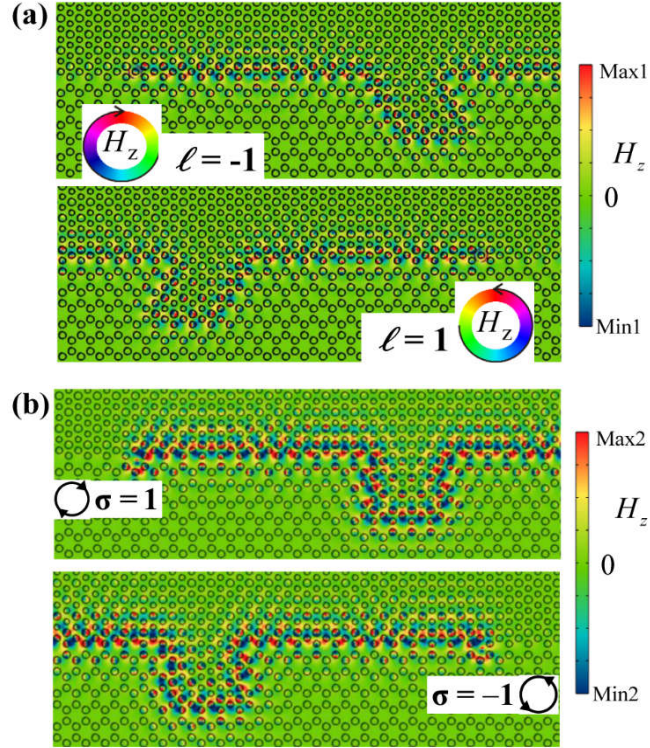

**Figure 3.** Robust and unidirectional edge states excited by (a) transverse orbital angular momentum and (b) transverse spin angular momentum sources, with the working frequency of 0.823 $c/a$. (a) The electromagnetic wave propagates towards right side and smoothly circumvents a $60°$ sharp bend with $l=-1$ ring source excitation, while it goes leftwards with $l=1$ ring source excitation. Here, the ring source has twenty-four equal-amplitude point sources on a circle with linear phase interval of $\pi/12$. The semi-diameter of the ring source is $0.6b$, and the ring center is at the white-arrow-surrounding rod in figure 2(c). (b) Selective excitation of robust unidirectional flow by the excitation of transverse spin angular momentum source. Pseudo-spin up

edge state (propagating rightwards) is excited by a RCP ($\sigma = 1$) source, while pseudo-spin down state (propagating leftwards) by a LCP ($\sigma = -1$) source. Note that the light flow circumvents the $90°$ sharp bend, manifesting the diverse design advantage of robust unidirectional flow in the future chip-size device.

Next, we will discuss the transmission spectra of the robust unidirectional flow when we change the size of ring source and the source position. It is expected that both cases may influence the selective excitation of robust unidirectional edge state as the local transverse angular momentum is position dependent. The transmission spectra of robust unidirectional flow along the $60°$ sharp bend is shown in figure 4(a), with same $l = -1$ source but different $R_s$ of the ring sources. The yellow background denotes the complete common gap and the green solid rectangle is the minigap. The transmission decreases little with $R_s$ value from 0.4$b$ (blue) to 0.6$b$ (green), but still keeps high level in broadband region, illustrating the predominance of strong photonic spin-orbit interaction. The little decrease can be understood by comparing the phase gradient of the TOAM source and the eigenmode. When $R_s = 0.4b$, the phase gradient of source is consistent with the eigenmode phase, resulting in the most mode match and thus the highest transmission in the complete gap except a narrow dip in the minigap [figure 4(a)]. However, when $R_s$ increases to 0.5$b$, a small part of ring source with clockwise increasing phase gradient will enter into a dashed region where the eigenmode phase gradient is inversely anti-clockwise. This mode mismatch leads to the slight decrease in transmission spectra. With $R_s$ further increasing to 0.6$b$, the more mode mismatch occurs, the further transmission decreases. Figure 5 illustrates the case when the source position has a finite lateral shift while keeping $R_s = 0.5b$ unchanged. The nearly same transmission of the two cases indicates that the 0.1$b$ shift does not affect the selective excitation, as the phase gradient of the source along the circle still matches to the phase gradient in the eigenmode, see detail in figure 5(b).

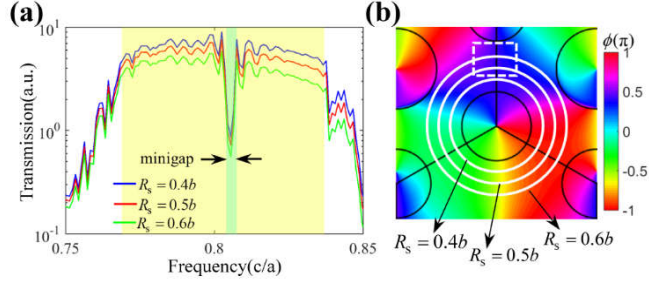

**Figure 4.** The influence of the size of ring size on the robust unidirectional flow along the edge with a 60° sharp bend. (a) Transmission spectra excited by a transverse orbital angular momentum source ($l=-1$) with different ring source size of $R_s = 0.4b$ (blue), $0.5b$ (red) and $0.6b$ (green), respectively. The yellow region represents the complete common gap and the pale green region represents the minigap. When $R_s$ is tuned from $0.4b$ to $0.6b$, the spectra decreases in the yellow region due to TAM mode mismatch. (b) Phase distribution of $H_z$ eigenfield in pseudo-spin up edge state at the frequency of 0.823 $c/a$. The radius of three white circles from inner to outer is $0.4b$, $0.5b$ and $0.6b$. The dashed rectangle indicates a region where the $H_z$ phase increases anti-clockwise along the white circle, opposite to the phase gradient of $l=-1$ source.

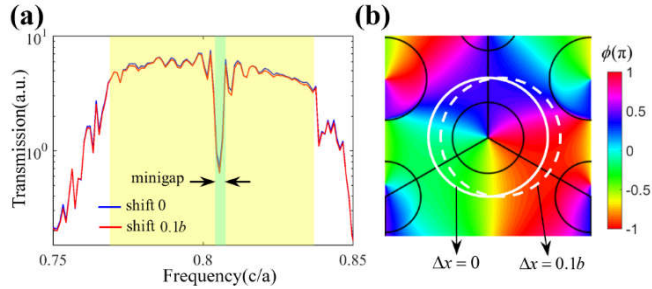

**Figure 5.** The influence on the same structure and same TOAM source ($l=-1$) as those of figure 4, except changing the source center location but fixing $R_s = 0.5b$. (a) Transmission spectra with different lateral shift of $\Delta x = 0$ (blue) and $\Delta x = 0.1b$ (red). Note that the two spectra are almost the same in the gap (yellow) region. (b) Phase distribution of $H_z$ eigenfield in pseudo-spin up edge state at the frequency of 0.823 $c/a$. The solid and dashed circles correspond to a lateral shift of 0 and $0.1b$, respectively.

## 4. Conclusion

We revealed the existence of transverse spin and orbital angular momenta in silicon topological photonic crystals by considering TE polarization. Polarization ellipse and phase distribution of pseudo-spin states shows their intrinsic transverse spin and orbital angular momenta. Besides, we found a one-dimensional vortex chain with alternating-sign topological charge at the boundary of topological photonic crystals. Taking advantage of transverse spin or orbital angular momentum mode with different chirality as the pseudo-spin source, we demonstrated the selective excitation of robust unidirectional light flow along several kinds of sharp bend, manifesting the diverse design advantage of robust unidirectional light flow in the future chip-size device. Moreover, we found that the robustness of unidirectional light flow can be maintained even when the ring-size or position of source varies in a certain range. Our work paves the way to realize robust unidirectional light flow in telecom region and may bring some potential applications in integrated photonic circuits such as on-chip robust delay line. Revealing the local transverse angular momentum in topological photonic crystals may also promote the study on the novel phenomenon in classical and quantum optics, especially in the field of chiral quantum optics.

**Acknowledgments**

This work is supported by Natural Science Foundation of China (11522437, 61775243, 11704422), Guangdong Natural Science Funds for Distinguished Young Scholar (S2013050015694) and Guangdong special support program.

**References**

[1] Grigorenko A N, Roberts N W, Dickinson M R and Zhang Y 2008 Nanometric optical tweezers based on nanostructured substrates *Nat. Photon.* **2** 365-70
[2] Eigler D M and Schweizer E K 2000 Positioning single atoms with a scanning tunnelling microscope *Nature* **344** 524-26
[3] Mair A, Vaziri A, Weihs G and Zeilinger A 2001 Entanglement of the orbital angular momentum states of photons *Nature* **412** 313-16
[4] Dong J-W, Chen X-D, Zhu H, Wang Y and Zhang X 2017 Valley photonic crystals for control of spin and topology *Nat. Mater.* **16** 298-302
[5] Chen X-D, Zhao F-L, Chen M and Dong J-W 2017 Valley-contrasting physics in all-dielectric photonic

crystals: Orbital angular momentum and topological propagation *Phys. Rev.* B **96** 020202(R)-06(R)

[6] Sollner I, et al. 2015 Deterministic photon-emitter coupling in chiral photonic circuits *Nat. Nanotechnol.* **10** 775-78

[7] Young A B, Thijssen A C, Beggs D M, Androvitsaneas P, Kuipers L, Rarity J G, Hughes S and Oulton R 2015 Polarization engineering in photonic crystal waveguides for spin-photon entanglers *Phys. Rev. Lett.* **115** 153901-05

[8] Lodahl P, Mahmoodian S, Stobbe S, Rauschenbeutel A, Schneeweiss P, Volz J, Pichler H and Zoller P 2017 Chiral quantum optics *Nature* **541** 473-80

[9] Wang Z, Chong Y, Joannopoulos J D and Soljačić M 2008 Reflection-free one-way edge modes in a gyromagnetic photonic crystal *Phys. Rev. Lett.* **100** 013905-08

[10] Poo Y, Wu R-x, Lin Z, Yang Y and Chan C T 2011 Experimental realization of self-guiding unidirectional electromagnetic edge states *Phys. Rev. Lett.* **106** 093903-07

[11] Wang Z, Chong Y, Joannopoulos J and Soljačić M 2009 Observation of unidirectional backscattering-immune topological electromagnetic states *Nature* **461** 772-75

[12] Fang K, Yu Z and Fan S 2012 Realizing effective magnetic field for photons by controlling the phase of dynamic modulation *Nat. Photon.* **6** 782-87

[13] Khanikaev A B, Mousavi S H, Tse W-K, Kargarian M, MacDonald A H and Shvets G 2013 Photonic topological insulators *Nat. Mater.* **12** 233-39

[14] Chen W-J, Jiang S-J, Chen X-D, Zhu B, Zhou L, Dong J-W and Chan C T 2014 Experimental realization of photonic topological insulator in a uniaxial metacrystal waveguide *Nat. Commun.* **5** 5782-88

[15] Wu L-H and Hu X 2015 Scheme for achieving a topological photonic crystal by using dielectric material *Phys. Rev. Lett.* **114** 223901-05

[16] Xiao B, Lai K, Yu Y, Ma T, Shvets G and Anlage S M 2016 Exciting reflectionless unidirectional edge modes in a reciprocal photonic topological insulator medium *Phys. Rev.* B **94** 195427-31

[17] Yang Y, Xu Y, Xu T, Wang H, Jiang J, Hu X and Hang Z 2016 Visualization of unidirectional optical waveguide using topological photonic crystals made of dielectric material *arXiv* 1610.07780

[18] Hafezi M, Mittal S, Fan J, Migdall A and Taylor J M 2013 Imaging topological edge states in silicon photonics *Nat. Photon.* **7** 1001-05

[19] Rechtsman M C, Zeuner J M, Plotnik Y, Lumer Y, Podolsky D, Dreisow F, Nolte S, Segev M and Szameit A 2013 Photonic Floquet topological insulators *Nature* **496** 196-200

[20] Tan W, Sun Y, Chen H and Shen S Q 2014 Photonic simulation of topological excitations in metamaterials *Sci. Rep.* **4** 3842-48

[21] Chen X-D, Deng Z-L, Chen W-J, Wang J-R and Dong J-W 2015 Manipulating pseudospin-polarized state of light in dispersion-immune photonic topological metacrystals *Phys. Rev.* B **92** 014210-17

[22] Gao W, Lawrence M, Yang B, Liu F, Fang F, Beri B, Li J and Zhang S 2015 Topological photonic phase in chiral hyperbolic metamaterials *Phys. Rev. Lett.* **114** 037402-06

[23] Lu L, Wang Z, Ye D, Ran L, Fu L, Joannopoulos J D and Soljačić M 2015 Experimental observation of Weyl points *Science* **349** 622-24

[24] Cheng X, Jouvaud C, Ni X, Mousavi S H, Genack A Z and Khanikaev A B 2016 Robust reconfigurable electromagnetic pathways within a photonic topological insulator *Nat. Mater.* **15** 542-48

[25] Gao F, et al. 2016 Probing topological protection using a designer surface plasmon structure *Nat. Commun.* **7** 11619-27

[26] He C, Sun X, Liu X, Lu M, Chen Y, Feng L and Chen Y 2016 Photonic topological insulator with broken

time-reversal symmetry *Proc. Natl. Acad. Sci.* **113** 4924-28

[27] Lu L, Fang C, Fu L, Johnson S G, Joannopoulos J D and Soljačić M 2016 Symmetry-protected topological photonic crystal in three dimensions *Nat. Phys.* **12** 337-40

[28] Shalaev M I, Desnavi S, Walasik W and Litchinitser N M 2017 Reconfigurable topological photonic crystal *arXiv* 1706.05325

[29] Weimann S, Kremer M, Plotnik Y, Lumer Y, Nolte S, Makris K G, Segev M, Rechtsman M C and Szameit A 2017 Topologically protected bound states in photonic parity-time-symmetric crystals *Nat. Mater.* **16** 433-38

[30] Mildred S. Dresselhaus, Gene Dresselhaus and Jorio A 2008 *Group Theory: Application to the Physics of Condensed Matter* ed Adelheid Duhm (Berlin: Springer-Verlag Berlin Heidelberg) pp 316-319

[31] Bernevig B A, Hughes T L and Zhang S-C 2006 Quantum spin Hall effect and topological phase transition in HgTe quantum wells *Science* **314** 1757-61

[32] Shen S-Q 2012 *Topological insulators: Dirac equation in condensed matters (Springer Series in Solid-State Sciences* vol 174*)* ed M. Cardona, et al (Berlin; New York: Springer-Verlag Berlin Heidelberg) p 23

[33] Mei J, Chen Z and Wu Y 2016 Pseudo-time-reversal symmetry and topological edge states in two-dimensional acoustic crystals *Sci. Rep.* **6** 32752-58

[34] Allen L, Beijersbergen M W, Spreeuw R and Woerdman J 1992 Orbital angular momentum of light and the transformation of Laguerre-Gaussian laser modes *Phys. Rev.* A **45** 8185-89

[35] Lu J, Qiu C, Ke M and Liu Z 2016 Valley vortex states in sonic crystals *Phys. Rev. Lett.* **116** 093901-06

[36] Leach J, Yao E and Padgett M J 2004 Observation of the vortex structure of a non-integer vortex beam *New. J. Phys.* **6** 71